\documentstyle{amsppt}\TagsOnRight\nologo
\newcount\refcount
\advance\refcount 1
\def\newref#1{\xdef#1{\the\refcount}\advance\refcount 1}
\newref\Bennettetal
\newref\superoperator
\newref\RainsEPP
\newref\VedralPlenio
\newref\sepsup
\newref\RainsPPT
\newref\Peres
\newref\HorodeckiPT
\newref\HHH
\newref\DShorSmolin
\newref\sentinel

\let\cal=\Cal

\topmatter
\title A rigorous treatment of distillable entanglement \endtitle
\author 
E. M. Rains
\endauthor
\affil AT\&T Research \endaffil
\address AT\&T Research, Room C290, 180 Park Ave.,
         Florham Park, NJ 07932-0971, USA \endaddress
\email rains\@research.att.com \endemail
\date October 12, 1998\enddate
\abstract
The notion of distillable entanglement is one of the fundamental concepts
of quantum information theory.  Unfortunately, there is an apparent
mismatch between the intuitive and rigorous definitions of distillable
entanglement.  To be precise, the existing rigorous definitions impose
the constraint that the distilation protocol produce an output of constant
dimension.  It is therefore conceivable that this unnecessary constraint
might have led to underestimation of the true distillable entanglement.
We give a new definition of distillable entanglement which removes
this constraint, but could conceivably overestimate the true value.
Since the definitions turn out to be equivalent, neither underestimation
nor overestimation is possible, and both definitions are arguably correct.

\endabstract
\endtopmatter
Since the concept of distillable entanglement is such a fundamental part
of quantum information theory, it is unfortunate that a gap currently
exists between its intuitive and rigorous definitions.

Intuitively, the distillable entanglement of a state $\rho$ is the maximum
over all allowable protocols of the expected rate at which ``good'' EPR
pairs can be obtained from a sequence of identical states.  For instance,
if we have a protocol which, given 10 copies of a state $\rho$, produces 10
``good'' EPR pairs half the time, and fails the other half, then we would
consider the distillable entanglement of $\rho$ to be at least $1/2$.
Unfortunately, it is not entirely obvious how to make this rigorous; in
particular, how one should take into account imperfect output when the
output dimension can vary.  For this reason, rigorous definitions
\cite\Bennettetal\ of distillable entanglement have so far only permitted
protocols which always produce the same sort of output; by these
definitions, we would only be justified in claiming that $\rho$ has
distillable entanglement at least $1/2$ if the above protocol produced 5
``good'' EPR pairs all the time, rather than 10 only half the time.
Consequently, these definitions could conceivably have underestimated the
``true'' distillable entanglement.

The purpose of the present note is to argue that this is not the case,
by giving two new rigorous definitions of distillable entanglement
which arguably would {\it overestimate} the intuitive distillable
entanglement, and then showing that the new definitions agree with
the existing definitions.

\head Classes of superoperators\endhead

The concept of distillable entanglement is not quite intrinsic to a state;
rather, the distillable entanglement can only be defined relative to some
specified class of legal operations.  It will be necessary, therefore, for
us to describe which such classes we will be considering.

Recall that any physical operation can be described by a ``completely
positive trace-preserving superoperator'' \cite\superoperator, that is an
operator ${\cal S}$ acting linearly on Hermitian matrices such that
$1\otimes {\cal S}$ takes density operators to density operators.  Any such
operator can be written in the form
$$
\rho \mapsto \sum_i S_i \rho S_i^\dagger = {\cal S}(\rho),
$$
with
$$
\sum_i S_i^\dagger S_i = 1.
$$
In practice, it is helpful to allow operations which are partially
classical; that is ``measurements''.  This corresponds to a decomposition
of ${\cal S}$ as a sum $\sum_i {\cal S}_i$ in which each ${\cal S}_i$ is a
completely positive, but not trace-preserving, superoperator mapping to a
Hilbert space $V_i$.  To be precise, each ${\cal S}_i$ is of the form
$$
\rho \mapsto \sum_j S_{ij} \rho S_{ij}^\dagger
$$
where each $S_{ij}$ has image in $V_i$, satisfying the condition
$$
\sum_i \sum_j S^\dagger_{ij} S_{ij} = 1.
$$
The key points are that the spaces $V_i$ need not be
the same, and that the operation also produces classical information
indicating which ${\cal S}_i$ is applied.  These will be the basic
operations allowed in the sequel, and will be referred to simply as
``operations''.  An operation consisting of more than one superoperator
will be said to be ``measuring''.

There is a natural notion of composition on operations; given an operation
${\cal S}$ on a Hilbert space $V$ and an operation ${\cal T}$ on the output
space $V_i$ of ${\cal S}$, one can compose ${\cal S}$ and ${\cal T}$ in the
obvious way (perform ${\cal S}$; if ${\cal S}_i$ was performed, then
perform ${\cal T}$).  One can also take tensor products of operations; if
${\cal S}=\{{\cal S}_i\}$ and ${\cal T}=\{{\cal T}_j\}$, then we define
$$
{\cal S}\otimes {\cal T} = \{{\cal S}_i\otimes {\cal T}_j\}.
$$
Finally, if ${\cal S}$ is an operation such that ${\cal S}_i$ and ${\cal
S}_j$ have the same output space, one can produce a new operation that
``forgets'' which of $i$ or $j$ occurred.

\proclaim{Definition} A ``class'' of operations is a set of operations
containing the identity and closed under all of the transformations of the
above paragraph.
\endproclaim

On a bipartite Hilbert space $V_A\otimes V_B$, there are five natural
classes that have been considered in the literature:

\item{$\bullet$} Local operations.  This is the class of operations of the
form
$$
{\cal S}_A\otimes {\cal S}_B,
$$
where ${\cal S}_A$ is a non-measuring operation on $V_A$ and ${\cal S}_B$
is a non-measuring operation on $V_B$.

\item{$\bullet$} 1-local operations (local operations plus one-way classical
communication).  This is the class generated by the local operations
together with all operations of the form
$$
{\cal S}_A\otimes 1,
$$
where ${\cal S}_A$ is an arbitrary operation on $V_A$.
(Here, the classical communication is from $A$ to $B$)

\item{$\bullet$} 2-local operations.  This is the class generated by
local operations,
$$
{\cal S}_A\otimes 1
$$
for operations ${\cal S}_A$, and
$$
1\otimes {\cal S}_B
$$
for operations ${\cal S}_B$.

\item{$\bullet$} Separable operations.  \cite{\RainsEPP,\VedralPlenio,\sepsup}\
This is the class of operations ${\cal S}$ such that each suboperation
${\cal S}_i$ of ${\cal S}$ is separable; that is, we can write
$$
{\cal S}_i(\rho) = \sum_j (A_j\otimes B_j) \rho (A_j\otimes B_j)^\dagger
$$
for operators $A_j$ and $B_j$ on $V_A$ and $V_B$ respectively.

\item{$\bullet$} Positive-partial-transpose (p.p.t.) operations.
\cite\RainsPPT\ This is the class of operations ${\cal S}$ such that each
suboperation ${\cal S}_i$ has completely positive partial transpose; that
is, the superoperator
$$
{\cal S}_i^\Gamma: \rho\mapsto {\cal S}_i(\rho^\Gamma))^\Gamma
$$
is completely positive, where $\Gamma$ is the partial transpose \cite\Peres.

The first three are classes by definition, and the last two are easily
verified to be classes.  It is also not too hard to verify that each class
in our list is contained in the next.  In fact, the containment is strict
in each case.  An example of a separable but not 2-local operation is given
in \cite\sepsup, while the creation of a p.p.t. but not separable state
(see, e.g. \cite\HorodeckiPT) is an inseparable, but p.p.t., operation;
the other cases are trivial.

\head Definitions of distillable entanglement\endhead

Associated to any class $C$ containing the class of local operations is a
notion of distillable entanglement.  As we have said, the $C$-distillable
entanglement of a state $\rho$ is intuitively defined as the rate at which
``good'' EPR pairs can be produced from copies of $\rho$ using only
operations from $C$.  However, as stated this is not a rigorous definition.

The prototype of our definitions of distillable entanglement is

\proclaim{``Definition''} The $C$-distillable entanglement of a state
$\rho$ on $V_A\otimes V_B$ is the maximum number $D_C(\rho)$ such that
there exists a sequence of operations ${\cal T}_i$ from $C$, where ${\cal
T}_i$ takes input $(V_A\otimes V_B)^{\otimes n_i}$, with outputs of the
form $V_{ij}\otimes V_{ij}$, and such that, as $i$ tends to $\infty$, we
have the limits $n_i\to\infty$,
$$
{1\over n_i} \sum_j p_{ij} \log_2 \dim V_{ij} \to D_C(\rho),
$$
and the output of ${\cal T}_i(\rho^{\otimes n})$ is ``good''.
Here $p_{ij}$ is the probability that the suboperation ${\cal
T}_{ij}$ is performed given the input state $\rho^{\otimes n_i}$.
\endproclaim

To define ``good'', we will use the notion of fidelity.  To any Hilbert
space $V$ with chosen basis $v_i$, we can associate a maximally entangled
state
$$
\Phi^+(V) = {1\over \sqrt{\dim V}} \sum_i v_i\otimes v_i.
$$
This choice of maximally entangled state is by no means canonical; however,
since any two maximally entangled states of the same dimension are
equivalent under local unitary operators, the definitions below do not
depend on the particular choice of $\Phi^+(V)$.  Given this convention, the
fidelity of a state $\rho$ on $V\otimes V$ is defined by
$$
F(\rho) = \Phi^+(V) \rho \Phi^+(V)^\dagger.
$$
Associated to any ${\cal T}_i$ from our prototypical definition, then,
is the sequence of fidelities $F_{ij}$ of ${\cal T}_{ij}(\rho^{\otimes n})$.
Our main claim, then, is that if we insist that the notion of ``good''
should depend only on the sequences $F_{ij}$ and $\dim V_{ij}$, and
the class $C$ contains the class of 1-local operations, then there is
a unique notion of $C$-distillable entanglement.

One definition given in the literature \cite\Bennettetal\ is

\proclaim{Definition 1}
The $C$-distillable entanglement of a state $\rho$ on $V_A\otimes V_B$ is
the maximum number $D_C(\rho)$ such that there exists a sequence of
non-measuring operations ${\cal T}_i$ from $C$, where ${\cal T}_i$ takes
input $(V_A\otimes V_B)^{\otimes n_i}$, with output of the form $V_i\otimes
V_i$, and such that as $i$ tends to $\infty$, we have the limits
$n_i\to\infty$,
$$
{1\over n_i} \log_2 \dim V_i \to D_C(\rho)
$$
and
$$
F_i\to 1.
$$
\endproclaim

Strictly speaking, they made the additional restriction that $\dim V_i$
should be a power of 2 for all $i$; we will call the resulting definition
definition 1'.  However, by Theorem 2 below, these definitions are equivalent.

It is clear that a sequence of operations satisfying definition 1' can
indeed be said to have distilled ``good'' EPR pairs.  However, this
definition is stronger than one would like, intuitively; as evidence of
this, note that the ``recurrence'' protocol of \cite\Bennettetal\ does not
directly meet this definition.  The correct definition, therefore, should
allow the operations ${\cal T}_i$ to be measuring.  The problem here is
that it is not immediately clear how the condition $F_i\to 1$ should be
generalized.

One possibility is as follows.  If a protocol distills entanglement
at a given rate, then it should certainly be the case that the entanglement
of formation of the output of the protocol increases at that rate.
If we define $E_f(F,K)$ to be the minimum entanglement of formation of
a bipartite state of dimension $K\times K$ and fidelity $F$, then
this suggests

\proclaim{Definition 2} The $C$-distillable entanglement of a state
$\rho$ on $V_A\otimes V_B$ is the maximum number $D_C(\rho)$ such that
there exists a sequence of operations ${\cal T}_i$ from $C$, where ${\cal
T}_i$ takes input $(V_A\otimes V_B)^{\otimes n_i}$, with outputs of the
form $V_{ij}\otimes V_{ij}$, and such that as $i$ tends to $\infty$, we
have the limits $n_i\to\infty$,
$$
{1\over n_i} \sum_j p_{ij} \log_2 \dim V_{ij} \to D_C(\rho),
$$
and
$$
{1\over n_i} \sum_j p_{ij} E_f(F_{ij},\dim V_{ij}) \to D_C(\rho).
$$
\endproclaim

Remark. Equivalently, the fidelity condition can be stated as
$$
{1\over n_i} \sum_j p_{ij} (\log_2\dim V_{ij}-E_f(F_{ij},\dim V_{ij}))\to 0.
$$

One possible objection to definition 2 is that it does not seem to allow
the possibility of protocols which sometimes fail to produce any result.
This is only apparently a problem; failure can be modelled by the
production of a state of dimension 1 (and thus fidelity 1 and entanglement
of formation 0).

Definition 2, if anything, has the problem of being too weak, since
entanglement of formation is a rather large measure of entanglement.
Since this definition is equivalent to the too-strong definition 1 (by
theorem 3 below), this argues that this is, indeed, the ``right''
notion of distillable entanglement.

In practice, definition 2 is difficult to work with; it will be convenient,
therefore, to introduce yet another definition,

\proclaim{Definition 2'}
The $C$-distillable entanglement of a state $\rho$ on $V_A\otimes V_B$ is
the maximum number $D_C(\rho)$ such that there exists a sequence of
operations ${\cal T}_i$ from $C$, where ${\cal T}_i$ takes input
$(V_A\otimes V_B)^{\otimes n_i}$, with outputs of the form $V_{ij}\otimes
V_{ij}$, and such that as $i$ tends to $\infty$, $n_i\to\infty$, we have
the limits
$$
{1\over n_i} \sum_j p_{ij} \log_2 \dim V_{ij} \to D_C(\rho),
$$
and
$$
{1\over n_i} \sum_j p_{ij} (1-F_{ij})\log_2\dim V_{ij}
\to 0.
$$
\endproclaim

\proclaim{Theorem 1}
Definitions 2 and 2' are equivalent for all classes $C$.
\endproclaim

\demo{Proof}
To show this, we need to know how $E_f(F,K)$ behaves for $K$ large.
Although we have defined $E_f(F,K)$ by minimizing over all states of fidelity
$F$, it is clear by symmetry and the convexity of $E_f$ that
this minimum is attained by states of the form
$$
a \Phi^+(K) \Phi^+(K)^\dagger + b;
$$
we will call such a state an isotropic state of dimension $K$.
The theorem, then, follows from Lemma 1 following.
\qed\enddemo

\proclaim{Lemma 1}
The entanglement of formation $E$ of an isotropic state of dimension $K$
and fidelity $F$ satisfies
$$
F \log_2 K-H_2(F)\le E\le F \log_2 K,
$$
where
$$
H_2(F) = -F\log_2 F-(1-F)\log_2(1-F).
$$
\endproclaim

\demo{Proof}
For the upper bound, write the state as a convex combination of the
isotropic state of fidelity 1 (with entanglement of formation $\log_2 K$)
and the (separable) isotropic state of fidelity ${1\over K}$ (with
entanglement of formation 0), and use the convexity of the entanglement of
formation to obtain an upper bound of
$$
{FK-1\over K-1}\log_2 K = F\log_2 K-{1-F\over K-1}\log_2 K \le F\log_2 K.
$$

For the lower bound, we use the fact \cite\RainsPPT\ that $E$ is bounded
below by the positive-partial-transpose bound on distillable entanglement.
For isotropic states, this bound was explicitly calculated to be
$$
\align
\log_2 K + F \log_2 F &+ (1-F)\log_2(1-F)-(1-F)\log_2(K-1)\\
&=
F \log_2 K-H_2(F)+(1-F)\log_2(K/(K-1))\\
&\ge
F \log_2 K-H_2(F).
\endalign
$$
\qed\enddemo

Another definition which has been proposed \cite\HHH\ replaces the
fidelity condition by
$$
\inf_j F_{ij} \to 1.
$$
Clearly, the distillable entanglement according to this definition lies
strictly between the values according to definitions 1 and 2', so
equivalence to our definitions follows from theorem 3 below.

\head Basic protocols\endhead

To show the remaining equivalences between the definitions, we will need
some basic transformations of isotropic states.  For instance, if we are
given an isotropic state of dimension $K$, to what extent can we transform
this into an isotropic state of dimension $K'<K$ without significantly
reducing the fidelity?  We consider two protocols, both local and symmetric
between ``Alice'' and ``Bob'' (the two subsystems).

In the first protocol, Alice's portion of the protocol is to measure the
subspace generated by the first $K'$ basis elements.  If Alice finds that
her portion of the state is in that subspace, she does nothing; otherwise,
she fails, i.e., replaces her portion of the state with a random element of
that subspace.  Bob performs the same protocol.

If Alice and Bob are given an isotropic state of fidelity 1, it is easy to
see that this protocol produces an isotropic state of fidelity 1 if both
Alice and Bob succeed in their measurements (probability $K'/K$), and otherwise
the protocol produces a completely random state.  On the other hand, on
a completely random state, the protocol will produce a completely random
state.  Thus the protocol must take the state
$$
a \Phi^+(K)\Phi^+(K)^\dagger + {1-a\over K^2}
$$
to the state
$$
{K'\over K}a\Phi^+(K')\Phi^+(K')^\dagger + {1-a {K'\over K}\over {K'}^2}.
$$
In other words, the state of fidelity $F$ is taken to the state of fidelity
$$
(K'/K) F + {(K-K')\left((1-F)K'(K'+K)+K^2-1\right)\over {K'}^2 K (K^2-1)}
\ge
(K'/K) F.
$$

In the second protocol, we require that $K'$ be a factor of $K$.  Both 
Alice and Bob interpret their state space as a tensor product of spaces
of dimension $K'$ and $K/K'$, then trace away the space of dimension $K/K'$.
Here a state of fidelity 1 maps to a state of fidelity 1, while a random
state maps to a random state.  Thus the state
$$
a \Phi^+(K)\Phi^+(K)^\dagger + {1-a\over K^2}
$$
is taken to the state
$$
a \Phi^+(K')\Phi^+(K')^\dagger + {1-a\over {K'}^2},
$$
or in other words, the state of fidelity $F$ is taken to the state
of fidelity
$$
f+(1-f){K^2-{K'}^2\over (K^2-1){K'}^2}\ge f.
$$

Combining these protocols, we obtain the lemma

\proclaim{Lemma 2}
For any pair $K'<K$, there exists a local operation which, given an
isotropic state of dimension $K$ and fidelity $F$, produces an
isotropic state of dimension $K'$ and fidelity at least
$$
{K'\over K} \lfloor{K\over K'}\rfloor F \ge {\max(K-K',K')\over K} F.
$$
More generally, for any state of dimension $K$ and fidelity $F$,
there exists a local operation which produces a state of dimension $K'$
and fidelity as stated.
\endproclaim

To be precise, we first use protocol 1 to reduce the dimension to
$K'\lfloor{K\over K'}\rfloor$ and then use protocol 2 to reduce the rest of
the way.  For non-isotropic states, we note that if we were to ``twirl''
\cite\Bennettetal\ the state by a random operator of the form $U\otimes
\overline{U}$, we would get an isotropic state of the same fidelity.
Since twirling is not local (only 1-local), this is not quite enough.
However, as in \cite\Bennettetal, one can then argue that {\it some} choice
of $U$ must obtain this fidelity, since the average $U$ does so, and
fidelity is linear.  So for $K'/K$ close to either 0 or 1, we can reduce to
dimension $K'$ without significantly reducing the fidelity, via purely
local operations.  We do not know what can be done in general for
intermediate values of $K'/K$.  (Locally, that is; if one allows 1-local
operations, one can simply teleport half of a maximally entangled state of
dimension $K'$ through the given state (M. and P. Horodecki, personal
communication).)  However, what we have shown is enough to give

\proclaim{Theorem 2}
Definitions 1 and 1' are equivalent for any class $C$ containing that
of local operations.
\endproclaim

\demo{Proof}
Clearly, any sequence of operations giving a lower bound on $D_C(\rho)$
according to definition 1' also satisfies the conditions of definition 1.
Suppose, therefore, that we are given a sequence of operations satisfying
the conditions of definition 1.  We need to show that there exists
a sequence of operations of the same rate in which the output always has
dimension a power of 2.

Let $K_i$ be the sequence of output dimensions.  Let $K'_i$ be defined for
each $i$ to be the largest power of 2 less than $K_i/n_i$.  Then we observe
the following:
$$
\align
\lim_{i\to\infty} {1\over n_i} \log_2 K'_i &=
\lim_{i\to\infty} {1\over n_i} \log_2 K_i.\\
\lim_{i\to\infty} K'_i/K_i &= 0.
\endalign
$$
In particular, applying lemma 2, we can produce a new sequence of operations
with output dimensions $K'_i$ and with output fidelities
$$
(1-{K'_i\over K_i}) F_i
$$
tending to 1.  Since the $K'_i$ are powers of 2, and have the same
value of ${1\over n_i} \log_2 K'_i$ as $i\to\infty$, we are done.
\qed\enddemo

Similarly, in definition 2' we may assume that all output dimensions are
powers of 2; the only complication is that some $K_{ij}$ might be less
than $n$, making $K'_{ij}$ less than 1.  This is simple to fix, however: if
$K_{ij}<n$, take $K'_{ij}=1$, making $F'_{ij}=1$.

To show that definitions 1 and 2' are equivalent, we will need the
following result:

\proclaim{Lemma 3}
If $K$ is a power of 2, then the 1-locally distillable entanglement
(according to definition 1) $D_1(F,K)$ of the isotropic state of fidelity
$F$ and dimension $K$ satisfies
$$
D_1(F,K)\ge (2F-1)\log_2 K -H_2(F)\ge (2F-1)\log_2 K-1.
$$
\endproclaim

\demo{Proof}
The ``hashing'' protocol \cite\Bennettetal, as extended in \cite\DShorSmolin,
gives
$$
D_1(F,K)\ge \log_2 K + F\log_2 F+(1-F) \log_2 ((1-F)/(K^2-1)).
$$
But
$$
\align
\log_2 K + F\log_2 F&+(1-F) \log_2 ((1-F)/(K^2-1))\\
&=
(2F-1)\log_2 K -H_2(F) + (1-F) \log_2 (K^2/(K^2-1))\\
&\ge
(2F-1)\log_2 K - H_2(F).
\endalign
$$
\qed\enddemo

\demo{Remark}
(1) Indeed, this is true when $K$ is a power of an arbitrary prime (H. Barnum,
D. DiVincenzo, personal communication), but we will not need this in
the sequel.  It is not clear what happens for general $K$.
(2) It is easy to verify that this is true for a state of dimension 1,
since then $F=1$, and in that case the bound says only that $D_1(1,1)\ge
0$.  (3) This result is true even if we insist in definition 1 that
$n_i=i$.  (4) It would be nice to have some bound of this sort be true
using only local operations; to be precise, if one could show that
$$
D_0(F,K)\ge (aF-(a-1))\log_2 K-o(\log_2 K)
$$
as $K\to\infty$, for some constant $a$, then this would allow theorem 3 to
be proved using only local operations.  Alternatively, if one could
show that $D_0=0$ for all impure states, the question of whether the
definitions are equivalent given only local operations would become moot.
\enddemo

\head The main theorem\endhead

\proclaim{Theorem 3}
If $C$ contains the class of 1-local operations, then definitions
1, 1', 2, and 2' are all equivalent.
\endproclaim

\demo{Proof}
By theorems 1 and 2, it suffices to show that definitions 1 and 2' are
equivalent.  Certainly, any sequence of operations satisfying the
conditions of definition 1 will also satisfy the conditions of definition
2'.  Suppose, therefore, that we are given a sequence of operations ${\cal
T}_i$ from $C$ satisfying the conditions of definition 2'.  Moreover,
assume that each output dimension is a power of 2 (which we can do,
by the remark after theorem 2).  Finally, by 1-local twirling, we may
insist that the output is always an isotropic state.

For each $i$, consider the operation ${\cal T}_i^{\otimes k}$ for large
$k$.  For any set of probabilities $p'_{ij}<p_{ij}$, the probability that
${\cal T}_i^{\otimes k}$ produces at least $\lfloor p'_{ij} k\rfloor$
copies of output $j$ can be made arbitrarily close to $1$ by taking $k$
sufficiently large.  If we also choose numbers $R'_{ij}$
with
$$
R'_{ij} < (2F_{ij}-1)\log_2 K_{ij}-1,
$$
then lemma 3 tells us that, given $p'_{ij} k$ states of dimension $K_{ij}$
and fidelity $F_{ij}$, we can produce, via 1-local operations, states of
dimension $\lfloor 2^{R'_{ij} p'_{ij} k}\rfloor$ with fidelity tending to 1 as
$k$ tends to infinity.

We can thus use the following protocol.  Apply ${\cal T}_i^{\otimes k}$
for sufficiently large $k$.  If we obtain at least $p'_{ij} k$
states of type $j$, then apply the hashing protocol to the states of type
$j$ for each $j$.  This results in a state of constant dimension
$$
K'_{i}(k)=\prod_j \lfloor 2^{R'_{ij} p'_{ij} k}\rfloor
$$
and fidelity tending to 1.  On the other hand, if we do not obtain the
desired numbers of states of each type, simply produce a random state
of dimension $K'$.  Since the probability of this occurring can be
made arbitrarily small, the resulting fidelity still tends to 1 as $k$
tends to infinity.  Thus we have a sequence of operations $O_k$ taking
as input $n_i k$ copies of $\rho$ and producing as output a state of
dimension $K'_i(k)$ with fidelity tending to 1.  This already tells
us that the $C$-distillable entanglement of $\rho$ according to
definition 1 is at least
$$
{1\over n_i} \sum_j R'_{ij} p'_{ij}.
$$
Since this is true for arbitrary $R'$ and $p'$ satisfying the above
inequalities, we have
$$
D_C(\rho) \ge ({1\over n_i} \sum_j p_{ij} (2F_{ij}-1) \log_2 K) - {1\over
n_i}
$$
for each $i$.  Letting $i$ tend to infinity, the theorem is proved.
\qed\enddemo

Remark.  A similar argument shows that we did not err in our definitions in
allowing an arbitrary sequence of input sizes $n_i$.  To be precise, for
any given rate $R$ less than the $C$-distillable entanglement, there is
certainly some $i$ such that the hashing protocol on the $i$th output
achieves rate at least $R$ asymptotically.  This gives a sequence of
operations with $n'_k = n_i k$.  But then for any number of inputs not in
this sequence, one can simply discard inputs as necessary, without
significantly changing the rate.  This gives a sequence of operations with
$n''_i=i$ demonstrating that $R\le D_C$.

\head Acknowledgements\endhead

The author is indebted to M. and P. Horodecki for pointing out the
problem with the existing definitions of distillable entanglement,
and for shooting down several early attempts at fixing the definitions.
The author would also like to thank D. DiVincenzo for helpful comments.

\enddocument
\bye